# Double-lens technique for efficient capture of short-lived particles by a crystal

V.M. Biryukov

Institute for High Energy Physics

Protvino, 142281, Russia

**Abstract.**

For the experiments to measure the dipole moments of short-lived particles via crystal channeling, we propose a beam-optics system of two crystal lenses that increases the number of channeled short-lived particles per one incident primary proton by a factor of about 1000 at LHC.

Particle beam channeling in bent crystals, pioneered in Dubna [1], was initially viewed as a low intensity application limited by a small Lindhard angle. This was indeed the case in the early years of bent crystal use for beam delivery at particle accelerators. Since the 80's crystals were exploited in external beam lines and in the ring of IHEP Protvino 70-GeV accelerator [2,3,4] on locations where traditional magnets could not do the required job.

To raise these applications to a new level, we did need the ideas how to increase the efficiency of crystals radically. At first glance, one cannot improve the technique because the critical angle of crystal lattice cannot be increased, and the typical GeV beam divergence is many orders of magnitude greater than it.

In 1990 it was theoretically found [5,6] that particle channeling in the accelerator rings should be based on a multipass mechanism. In a ring, the circulating particles may pass through the crystal many times. Despite of scattering and losses on every unsuccessful encounter with a crystal, the multiplicity of encounters increases the overall efficiency of channeling greatly. To benefit from multipass mode, the crystal had to be shortened strongly. Optimal curvature was found to be factor of 2-3 greater than for single-pass channeling in external beamlines. Although this strong curvature reduces the chances of channeling on every single pass, this is overcompensated with the increased mean number of passes in a short crystal.

This idea and its realization in IHEP [7-10] opened the way to high-efficiency and high-intensity applications of bent crystal technique for beam steering at accelerators. The new scheme of crystal extraction, based on multipass channeling, resulted in the increase of the intensity of extracted beam by 5 orders of magnitude, from $10^7$ to about $10^{12}$ proton per cycle [7]. The efficiency of crystal extraction of 70 GeV protons increased with introduction of short-crystal multipass channeling by 4 orders of magnitude, from 0.01% [2] to 85% [9]. Six locations on IHEP 70-GeV ring are equipped by crystal extraction systems, serving for routine applications [9].

The use of multipass mechanism was subsequently proposed for crystal extraction from LHC [11] and crystal collimation from the Tevatron [12], RHIC [13], and LHC [14]. While the first-generation crystals used in the rings were as long as 80 mm (IHEP [2]), 30 mm (SPS [15]), 40 mm (Tevatron [16]), later they were shortened to 2 mm (IHEP [9]), 4 mm (LHC [17]), and 5 mm in RHIC [18] and Tevatron [19] along the beam.

Any other ideas for radical increase of channeling efficiency would be of great help. One such idea we discuss below.

We shall consider the use of bent crystals for channeling of short-lived particles. This has raised a substantial interest in recent years with the purpose to measure magnetic dipole moments (MDM) and electric dipole moments (EDM) of charmed particles and tau leptons at LHC [20-24]. It was proposed to extract (by means of a channeling crystal) a proton beam from LHC ring, which subsequently hits a tungsten target and produces short-lived particles. These particles are trapped by another crystal and bent a substantial angle which results in a spin precession of channeled particles allowing to measure MDM and EDM.

While there is a wide interest to apply this technique at LHC, one of the problems to be solved is how to obtain the necessary intensity of primary protons at the target and respectively the sufficient intensity of short-lived particles trapped in crystal. The required protons intensity is a minor fraction of the protons that are continuously lost from the LHC ring and intercepted by the collimation system. However, in order to get (extract from the LHC ring) these protons you need to install in the halo a channeling crystal that traps protons and bends them onto the target. To get enough protons, this crystal needs to be pretty close to the circulating beam in the LHC, and this may worry the collimation experts.

Another great bottleneck is the low probability of short-lived particles capture by a crystal because the divergence of these particles is three orders of magnitude larger than the Lindhard angle of crystal. Respectively, we loose a factor of 1000 or so at the capture into channeling mode. Plus, some reasonable losses follow when particles are bent by a channeling crystal.

Finding a compromised location for an extraction crystal, where it can get enough intensity and still does not interfere in the work of the LHC collimation system, is not easy, and we do not discuss it here.

For the other bottleneck it looks that little can be done. The divergence of the particles is set by the physics of the particles production and decay, and by their energy. The critical Lindhard angle is defined by crystal lattice constants and particles energy. By replacing silicon crystal with a heavier crystal like germanium, we can increase the Lindhard angle by about 30% [25]. By cooling the crystal to cryogenic temperature we could increase the Lindhard angle by another 10% or so [25]. Cooling also helps the crystal transmission efficiency [25,26]. These options are discussed and evaluated elsewhere [20-24]. For example, ref [24] estimated the increase in crystal channeling efficiency from replacing silicon crystal with germanium by about 20%.

In this paper we take a different approach to the problem. For crystal channeling physicists it is common to check the divergence of the incoming beam with the critical angle of crystal lattice. We cannot change the divergence or the critical angle sizably.

Let us look at the problem in another way. For accelerator scientists it is common to speak of the beam emittance, i.e. the phase space $(x,x')$ area occupied by the beam particles, rather than beam divergence. Respectively, it is common to speak of the beamline acceptance in phase space rather than of its "critical angle". This is because when the beam is transferred through a beamline its divergence and its size may change to a great extent, e.g. in every quadrupole lens, while the beam emittance is invariant.

The acceptance of the crystal, i.e. the phase space area available for channeling, is about the Lindhard angle times the transverse size of the crystal. There is no point in having the crystal wider than the incoming beam. And we cannot increase the Lindhard angle, at least not greatly. So the crystal acceptance is fixed and remains quite small because of a small Lindhard angle.

The beam emittance is defined by the divergence of the produced particles times the transverse size of the beam source. The divergence cannot be changed. However, we can reduce the beam size strongly. If the beam particles emerge from a point-like source, beam emittance can be quite small despite of a big divergence.

If we reduce beam emittance to a sufficiently small value that matches crystal acceptance, then we could get this beam channeled efficiently, in principle. Now, after matching, physics allows efficient channeling, so we only need to engineer the things in a proper way.

In order to solve the problem we need, first, to focus the primary proton beam on the tungsten target, so the source of the secondary particles will be a point-like. This can be done with a focusing crystal upstream of the target [27]. The actual transverse size $\sigma$ of the beam at focus equals $\pm\theta_{L1}\cdot L_1$, limited by Lindhard angle $\theta_{L1}$ of primary protons and focal length $L_1$ of the lens (focusing crystal) upstream of the target.

$$\sigma = \theta_{L1} \cdot L_1 \qquad (1)$$

Then we need to match this point-like beam source with another focusing crystal that has its focal point right at the beam source. This second crystal lens does the reverse job capturing the particles emerging from the point-like source.

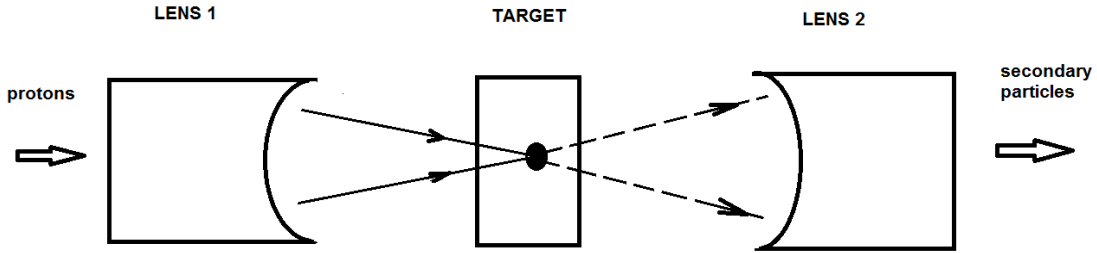

**Fig. 1** Schematic of two crystal lenses and a target between. First lens focuses primary protons on the target. Second lens traps secondary particles emerging from the target.

It was already experimentally demonstrated that a divergent beam emerging from a point-like source can be effectively captured by a focusing crystal [28]. In the IHEP Protvino experiment a beam of 70-GeV protons with divergence 100 times greater than Lindhard angle was channeled and deflected with efficiency of 15%, i.e. orders of magnitude greater than one would expect basing on simply the ratio of Lindhard angle to beam divergence.

In principle, we can even take any of the proposed schemes for crystal channeling of short-lived particles at LHC and leave it intact, but convert the plain crystal faces into focusing-like. In these schemes, the exit face of crystal 1 (the extraction crystal in the LHC halo, i.e. the crystal that sends protons to the target) should be made focusing with the focus at the target. The entry face of crystal 2 (that traps short-lived particles) should be made focusing with the focus at the target.

However, we believe it is much better to leave the extraction crystal intact, but make as a single assembly the two crystal lenses and the target as shown in Figure 1.

This assembly can be tuned with all its parts aligned to each other, offline. Then it would be much easier to use this pre-tuned assembly in the beam, instead of having to tune the parts of the scheme online. Also, it is essential to have the focal length of the first crystal lens (the one before the target) short enough in order to make the proton beam size at the target as small as possible.

The efficiency of a plain crystal strongly depends on a beam divergence. From the particles incident at crystal with divergence $\Phi$, only a fraction $\theta_L/\Phi$ of them fits the Lindhard angle $\theta_L$. Those that fit will suffer further losses from surface acceptance and dechanneling caused by bending and scattering in crystal. However, these further losses are moderate, so the major limitation on crystal channeling efficiency is the poor ratio $\theta_L/\Phi$ for divergent beams.

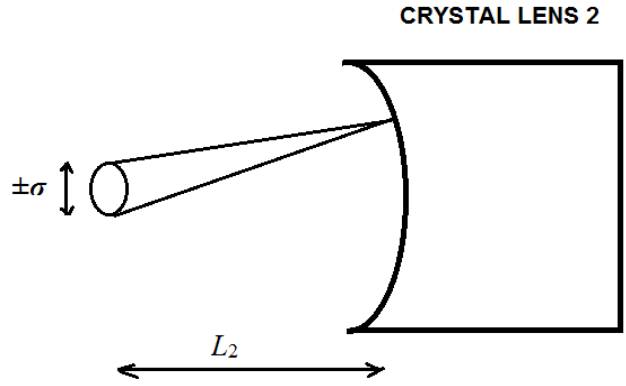

**Fig. 2** Beam source of size $\pm\sigma$ is seen by a crystal lens at a focal distance $L_2$ with effective divergence $\pm\sigma/L_2$ at any entrance point.

The efficiency of a focusing crystal is independent of the beam divergence. It depends on the transverse size $\sigma$ of the beam source at the focal point. At any given point on the entry face of the second crystal lens, the divergence $\Psi$ of the incoming particles is set by the size of the beam source $\sigma$ and the focal length $L_2$, see Figure 2

$$\Psi = \sigma/L_2 = \theta_{L1} \cdot (L_1/L_2) \qquad (2)$$

This value, $\Psi$, has to be compared with the critical angle $\theta_{L2}$ of the second crystal which has to channel the secondary particles. We see that, in the system of two lenses, the effective divergence of the secondary particles emerging from the focus and entering the crystal lattice depends on the Lindhard angle $\theta_{L1}$ of the primary protons and on the two focal lengths, $L_1$ and $L_2$.

Instead of ratio $\theta_{L2}/\Phi$ in the schemes with a plain crystal, in our scheme we have ratio $\theta_{L2}/\Psi$ for the probability of short-lived particles to be within Lindhard angle $\theta_{L2}$ at the second crystal downstream of the target. That is, our beam-optics idea gains in efficiency a factor $\Phi/\Psi$ over the traditional scheme

$$\Phi/\Psi = (\Phi/\theta_{L1}) \cdot (L_2/L_1) \qquad (3)$$

The number of channeled short-lived particles per one incident primary proton increases by the above factor $\Phi/\Psi$ compared to the classic plain-crystal scheme. The primary protons of 6.5 TeV have the Lindhard angle $\theta_{L1} \approx 2.2$ μrad in silicon (110) planes. We estimate the typical divergence of short-lived particles in the schemes of interest for particle physicists basing on ref. [22]. There, the transverse momenta $p_T$ of short-lived particles in the range of 1 to 3 GeV/c are of most interest as these particles have sufficient polarization. This implies a divergence of about $\Phi \approx 3$ mrad for the sample of short-lived particles at the crystal entrance.

Thus, the ratio $\Phi/\theta_{L1} \geq 1000$ gives a great advantage to our scheme. If we could make $L_1$ shorter than $L_2$, then we gain in Eq.(3) even more, like another factor of $L_2/L_1 \approx 2$ or so. However, when we reduce $L_1$, it reduces the efficiency of lens 1 while increases the efficiency of lens 2 at the same time. So the optimization over $L_1$ and $L_2$ should be done carefully taking into account all other aspects of the technique

The efficiency of the introduced first crystal lens, which focuses primary protons at the target, is not 100%, of course. Here we can roughly estimate it as follows. If the focal length $L_1$ is on the order of 0.1 m and the incident proton beam size is on the order of 1 mm, then the crystal lens bends the protons by 10 mrad at most, or 5 mrad on average. The efficiency of bending proton beam a few mrad is quite high, about 50% in the experiments [29]. So we expect to loose about a factor of 2 on the first lens.

If you make its focal length larger, 1 m, then the lens has to bend protons just 0.5 mrad, which can be done with greater efficiency. But the smaller focal length of first crystal is preferable because it makes the overall efficiency of two lenses system greater. The optimal focus length of the first lens should be found in analysis for the overall system, lens-target-lens sandwich, because it affects the efficiencies of both crystal lenses.

The theoretical limit, of the order of 1000, of this technique may be lowered in experiment by some practical factors. Lens optical aberration may cause broadening of the focus. However, in the experiments at Protvino in the 90's the achieved beam size at the focus agreed with theory, Lindhard angle times the focal length. We believe modern technology can handle it as well. Crystals with focal lengths below 0.5 m have been produced and tested in high-GeV beam [28].

This idea of two-lens objective promises a potential gain of a factor 1000 or so. In practice we need to work out every part of it in order to understand how much we can gain in reality. This double-lens technique needs a detailed simulation. It involves many variables, like the focusing lengths of both crystals, the decay lengths of short-lived particles, the distribution of the particles produced in the target (and in further decay) in some energy range etc. A thorough optimization of the technique based on realistic models is necessary. More important sophistication would come from the phase space distribution of particles produced in collision or decay. Several scenarios with different particles obtained via different ways were discussed [20-24].

This double-lens idea is richer than just that. If you want to trap the secondary particles emerging from the target, it is optimal that the focal points of the two lenses coincide (overlap at the target). But if you want to trap the decay products of the parent secondary (short-lived) particles that emerge from the target, then it is optimal to separate the two focal points of the lenses by the distance of about $\gamma c\tau$ which is the decay length of the parent short-lived particle with Lorentz factor $\gamma$ in the laboratory frame, as illustrated by Figure 3.

This is the case when a primary proton produces in the target a charm particle which decays shortly after and gives birth to a tau lepton which you want to channel in a crystal and study its dipole moments. This idea of focus separation from the target with the purpose of obtaining very clean beam of short-lived particles is presented in ref. [30] and discussed in some detail there. Notice that in this geometry with separated focus the second crystal traps solely the decay products (like tau lepton etc) while the debris from the target is rejected (with rejection factor of the order of $10^4$ [30]), thus providing clean samples of short-lived particles.

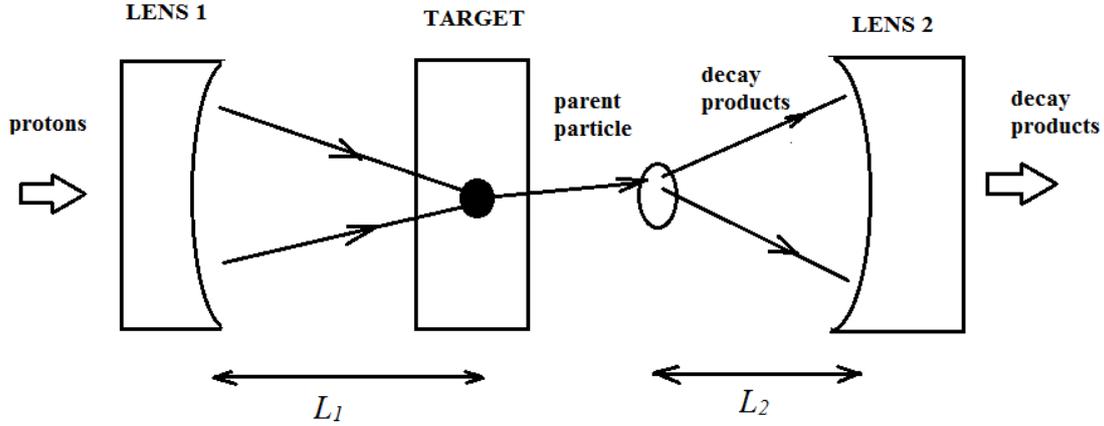

**Fig. 3** A variant of crystal lenses and target assembly. Here the second-lens focus is separated from the target to catch the decay products appearing downstream.

Although our purpose here is to present the idea, while the precise parameters should be defined later depending on specific particle physics needs, we suggest tentative characteristics of the lenses. Focal length $L_1 \approx$ 0.1 m looks reasonable. The tungsten target (of the same size as suggested elsewhere, e.g. 2 cm) is centered with the focus of both lenses. The distance $L_2$ between target and the second crystal (lens) should be the same as suggested in the earlier studies on this subject, on the order of $L_1$. The chosen distance $L_2$ should be the focal length of second crystal lens. The characteristics of this second crystal should be the same as suggested elsewhere for the studies of spin precession in it, but its entry face should be made focusing-like with a focus at the beam source. Channeling of primary 6.5-TeV protons in the second crystal is avoided because its curvature (e.g. 10 cm length, 16 mrad bend [24]) is stronger than Tsyganov threshold for 6.5 TeV while it is fine for 1-2 TeV particles.

In addition, together with the proposed technique, one can use other ideas discussed in the literature: replacing silicon with germanium, cooling the crystal to cryogenic temperature etc. These options do not contradict our scheme and can be used simultaneously with it.

Like the invention of the telescope and microscope gave us optical instruments much more powerful than a single lens, the present idea of two-lens objective could magnify the number of captured short-lived particles at LHC by a factor of about 1000 compared to a plain crystal. This enhancement in event rate would mean that we may need now much less time for data taking at LHC. Earlier, some physics ideas required years or decades of data taking. Now we could obtain the same statistics in just a few days. Some ideas were beyond reach because of unrealistically long data taking required. Now we could get back to these ideas.

Optimistic scenarios implied a few months [20-24] of running the experiment. Now we might complete the same experiment in a few hours. Earlier these experiments were assumed running in parasitic mode. Now, if an experiment could be done in a very short time, one might ask for a dedicated time.

Alternatively, you can view the 1000-fold increase in the number of channeled charm (tau) particles per one primary proton as the respective reduction in the need for the intensity of primary protons. This would greatly simplify the search for a suitable location of the extraction crystal in the halo of the LHC beam that satisfies both the collimation experts and the particle physicists.

# References

[1] A.F. Elishev et al. "Steering of the Charged Particle Trajectories by a Bent Crystal". Phys. Lett. B **88** (1979) 387

[2] A.A. Aseev et al, "Extraction of the 70-GeV proton beam from the IHEP accelerator towards beamline 2(14) with a bent single crystal". Nucl. Instr. Meth. A **309** (1991) 1

[3] A.A. Asseev et al, "The main results of four years experience on extraction of protons by bent crystal from the 70-GeV IHEP accelerator". Conf. Proc. C 940627 (1995) 2388

[4] V.M. Biryukov, V.I. Kotov, Y.A. Chesnokov. "Steering of high-energy charged particle beams by bent single crystals". Phys.Usp. **37** (1994) 937

[5] V. Biryukov, "On the theory of proton beam multiturn extraction with bent single crystal ". Nucl. Instrum. Meth. B **53** (1991) 202

[6] A.M. Taratin et al. "Computer simulation of multitum beam extraction from accelerators by bent crystals". Nucl. Instrum. Meth. B **58** (1991) 103

[7] A.G. Afonin et al., "High efficiency multipass extraction of 70-GeV protons from accelerator with a short bent crystal". Phys. Lett. **B 435** (1998) 240

[8] A.G. Afonin et al., "First results of experiments on high-efficiency single-crystal extraction of protons from the U-70 accelerator". JETP Lett. **67** (1998) 781

[9] A.G. Afonin et al., "The schemes of proton extraction from IHEP accelerator using bent crystals". Nucl. Instrum. Meth. **B 234** (2005) 14

[10] A.G. Afonin et al., "Proton beam extraction from the IHEP accelerator using short silicon crystals". Phys. Part. Nucl. **36** (2005) 21

[11] V. Biryukov. "Computer simulation of crystal extraction of protons from a Large Hadron Collider beam". Phys. Rev. Lett. **74** (1995) 2471

[12] V.M. Biryukov, A.I. Drozhdin, N.V. Mokhov. "On Possible use of bent crystal to improve Tevatron beam scraping". FERMILAB-CONF-99-072

[13] R.P. Fliller et al. "RHIC Crystal Collimation". Nucl. Instrum. Meth. **B 234** (2005) 47

[14] V.M. Biryukov et al. "Crystal collimation as an option for the large hadron colliders". Nucl. Instrum. Meth. **B 234** (2005) 23

[15] H. Akbari et al. "First results on proton extraction from the CERN SPS with a bent crystal". Phys. Lett. B **313**, 491 (1993)

[16] R.A. Carrigan Jr. et al., "Extraction from TeV range accelerators using bent crystal channeling". Nucl.Instrum.Meth. **B 90** (1994) 128

[17] W. Scandale et al. "Observation of channeling for 6500 GeV/ c protons in the crystal assisted collimation setup for LHC". Phys. Lett. **B 758** (2016) 129

[18] R.P. Fliller et al. "Results of bent crystal channeling and collimation at the Relativistic Heavy Ion Collider". Phys. Rev. ST Accel.Beams **9** (2006) 013501

[19] R. Carrigan et al. "Emittance Growth and Beam Loss". Chapter in book: V. Lebedev, V. Shiltsev (eds.) "Accelerator physics at the Tevatron Collider" (2014)

[20] V.G. Baryshevsky. "The possibility to measure the magnetic moments of short-lived particles (charm and beauty baryons) at LHC and FCC energies using the phenomenon of spin rotation in crystals". Phys. Lett. **B 757** (2016) 426

[21] E. Bagli et al. "Electromagnetic dipole moments of charged baryons with bent crystals at the LHC". Eur.Phys.J.C **77** (2017) 12, 828

[22] A.S. Fomin, A. Yu Korchin, A. Stocchi, S. Barsuk, P. Robbe. "Feasibility of tau - lepton electromagnetic dipole moments measurement using bent crystal at the LHC". JHEP **03** (2019) 156

[23] J. Fu, M.A. Giorgi, L. Henry, D. Marangotto, F. Martínez Vidal, A. Merli, N. Neri, and J. Ruiz Vidal. "Novel Method for the Direct Measurement of the tau Lepton Dipole Moments". Phys. Rev. Lett. **123** (2019) 011801

[24]  S. Aiola et al."Progress towards the first measurement of charm baryon dipole moments". Phys. Rev. D **103** (2021) 072003

[25]  V.M. Biryukov, Y.A. Chesnokov, V.I. Kotov. Crystal Channeling and Its Application at High-Energy Accelerators. 1997: Springer-Verlag Berlin and Heidelberg

[26]  J.S. Forster et al. "Deflection of GeV Particle Beams by Channeling in Bent Crystal Planes of Constant Curvature", Nucl. Phys. B **318** (1989) 301

[27]  M. A. Gordeeva et al., "First results on the focusing of a 70-GeV proton beam by a curved single crystal", JETP Lett. **54** (1991) 487

[28]  V. I. Baranov et al. "Highly efficient deflection of a divergent beam by a bent single crystal", Nucl. Instrum. Methods B **95** (1995) 449

[29]  A. Baurichter et al. "New results from the CERN SPS beam deflection experiments with bent crystals", Nucl. Instrum. Methods B **119** (1996) 172

[30]  V.M. Biryukov, "Possibility to make a beam of tau-leptons and charmed particles by a channeling crystal". ArXiv:2101.05085